\documentclass{PoS}

\usepackage{multirow}
\title{Light Glueball masses using the Multilevel Algorithm.}

\ShortTitle{Light Glueball masses using the Multilevel Algorithm.}

\author{\speaker{Sourav Mondal}\\
        Indian Association for the Cultivation of Science, Kolkata.\\
        E-mail: \email{tpsm5@iacs.res.in}}

\author{Pushan Majumdar\\
        Indian Assiciation for the Cultivation of Science, Kolkata. \\
        E-mail: \email{tppm@iacs.res.in}}
        
\author{Nilmani Mathur\\
        Tata Institute of Fundamental Research, Mumbai. \\
        E-mail: \email{nilmani@theory.tifr.res.in}}

\abstract{Following the multilevel scheme we present an error reduction algorithm for extracting glueball masses from monte-carlo
simulations of pure SU(3)
lattice gauge theory. We look at the two lightest states viz. the $0^{++}$
and $2^{++}$. Our method involves looking at correlations between large wilson loops and does not require any smearing of
links. The error bars we obtain are at the moment comparable to those obtained using smeared operators. We also present
a comparison of our method with the naive method.}

\FullConference{The 32nd International Symposium on Lattice Field Theory,\\
		23-28 June, 2014\\
		Columbia University New York, NY}

\begin{document}

\section{Introduction}

Low lying spectrum of pure Yang-Mills theory consists of glueballs. Quarks allow glueballs to decay but their 
signature is expected to remain in the QCD spectrum. While glueballs have not yet been discovered experimentally, 
there are candidates \cite{PDG}. For a recent review on the status of glueballs see
reference \cite{Ochs}. Glueball correlators can be computed in 
lattice gauge theory simulations but extraction of glueball masses from correlation functions
are extremely difficult because the correlation functions are dominated by statistical noise. Recent attempts 
to compute glueball masses in both pure Yang-Mills theory and lattice QCD using different strategies to combat 
the statistical noise have been reported in references \cite{HartTeper,Hart,Richards,teper,flow,Bali,Vaccarino,
Lucini,Morningstar,nilmani,meyer1,meyer2,meyer3,myerteper,DellaMorte}. In this article we 
report on our attempts to reduce the statistical noise on glueball correlators obtained from simulations of pure 
SU(3) Yang-Mills theory in 4 Euclidean dimensions. 

We explored the scalar ($0^{++}$) and the tensor ($2^{++}$) channels and 
to reduce the noise on the correlators in these channels, we tried the following two strategies 
(i) construct glueball operators from large wilson loops with extents of about half a fermi in each direction 
 and 
(ii) extract masses from the correlators with fit range as large as possible (0.5 $-$ 1.0 fermi) to reduce 
contamination from excited states. The idea to use large Wilson loops to construct glueball operators was 
proposed in \cite{Rgupta} but it was only recently that it was coupled to error reduction techniques to estimate the 
expectation values of the Wilson loops accurately \cite{PNS}. 

\section{Algorithm}

For updating the pure Yang-Mills fields we used the Cabibo-Marinari heatbath for $SU(3)$ with 3 over-relaxation 
steps for every heatbath step. Between each measurement we did 10 full sweeps of the lattice to ensure that 
successive measurements could be treated as independent.

The glueball operators were constructed using Wilson loops. 
For the scalar channel we constructed the temporal correlator between the operators 
$\left ( {\mathcal A} - \langle {\mathcal A}\rangle\right )$ at different time slices with
${\mathcal A}={\mathbb Re}\left ( P_{xy} + P_{xz} + P_{yz} \right )$ and for the tensor channel we took the
two operators ${\mathcal E}_1={\mathbb Re}\left ( P_{xz}
-P_{yz}\right )$ and ${\mathcal E}_2={\mathbb Re}\left ( P_{xz} + P_{yz} - 2 P_{xy} \right )$.
 Here $P_{ij}$ denotes a Wilson loop in the $ij$ plane with $ij$ going over the spatial directions. 

For the noise reduction scheme we used the philosophy of multilevel algorithm \cite{ml}.
This method is particularly useful in theories with a mass gap, where the distant regions
of the theory are uncorrelated as the correlation length is finite.

The principle of multilevel algorithm is to compute the expectation values in a nested manner.
Intermediate values are first constructed by averaging over sub-lattices
with boundaries and then the full expectation values are obtained by averging over the
intermediate values with different boundaries obtained by updating the full lattice.
The intermediate averages can be computed in a nested manner and our innermost 
 noise reduction step was to use a semi-analytic multihit on the SU(3) links \cite{deForcrand}
with which the Wilson loops were constructed. This reduced the fluctuations in the expectation 
values of the glueball operators. The multilevel, on top of the multihit, was used to reduce the 
fluctuation in the correlators.

\begin{figure}
\centerline{\includegraphics[width=0.6\textwidth,angle=0]{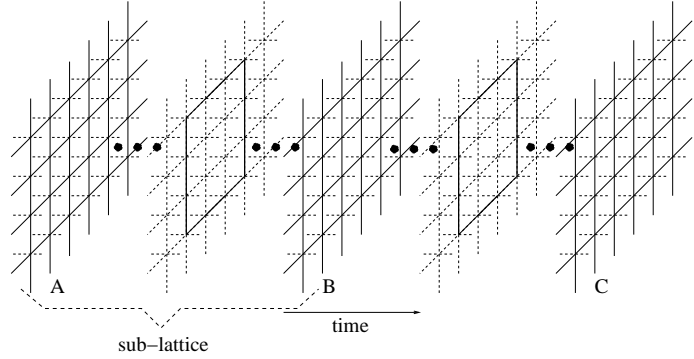}}
\caption{Slicing the lattice for the multilevel update. Solid links on slices A,B and C are frozen 
during the sub-lattice updates. Thick lines in the Wilson loops indicate that the corresponding 
links have been replaced by their multihit averages.} 
\label{fig:scheme}
\end{figure}

In figure \ref{fig:scheme} we illustrate our slicing of the lattice and the computation of the
intermediate expectation values of glueball operators (Wilson loops) by performing several sub-lattice 
updates.

The multilevel algorithm has a number of extra parameters such as the thickness of the sublattice and the 
number of sublattice updates used in the intermediate averages. Optimal values of these parametes depend on 
the observable one is trying to estimate and some tuning of these parameters are essential for efficient 
error reduction. In tables \ref{tab:param-scalar} and \ref{tab:param-tensor} we record these along with 
other simulation parameters for our runs. The scale is set through the Sommer parameter $r_0$ computed for these 
$\beta$ values in reference \cite{Sommer}.
 
\begin{table}
\centerline{
\begin{tabular}{|c|c|c|c|c|c|} \hline
Lattice Size & $\beta$ & $({r_0}/{a_t})$& $^{\rm sub-lattice}_{\rm thickness}$ & iupd & loop size \\ \hline
${10^3}\times18$ & 5.7 & 2.922(9) &3&30&$2\times 2$ \\ \hline
${12^3}\times18$ & 5.8 & 3.673(5) &3&25&$3\times 3$ \\ \hline
${16^3}\times24$ & 5.95 & 4.898(12) &4&50&$5\times 5$ \\ \hline
\end{tabular}}
\caption{Simulation parameters for the scalar channel}
\label{tab:param-scalar}
\end{table}

\begin{table}
\centerline{
\begin{tabular}{|c|c|c|c|c|c|} \hline
Lattice Size & $\beta$ & $({r_0}/{a_t})$& $^{\rm sub-lattice}_{\rm thickness}$ & iupd & loop size \\ \hline
${12^3}\times18$ & 5.8 & 3.673(5) &3&70&$3\times 3$ \\ \hline
${12^3}\times20$ & 5.95 & 4.898(12) &5&100&$5\times 5$ \\ \hline
${12^3}\times20$ & 6.07 & 6.033(17) &5&100&$5\times 5$ \\ \hline
\end{tabular}}
\caption{Simulation parameters for the tensor channel}
\label{tab:param-tensor}
\end{table}

\section{Results}
In our simulations we could follow the glueball correlators at least upto distances 
of about one fermi in the scalar channel and about 0.8 fermi in the tensor 
channel. 

\begin{figure}
\includegraphics[width=0.35\textwidth,angle=270]{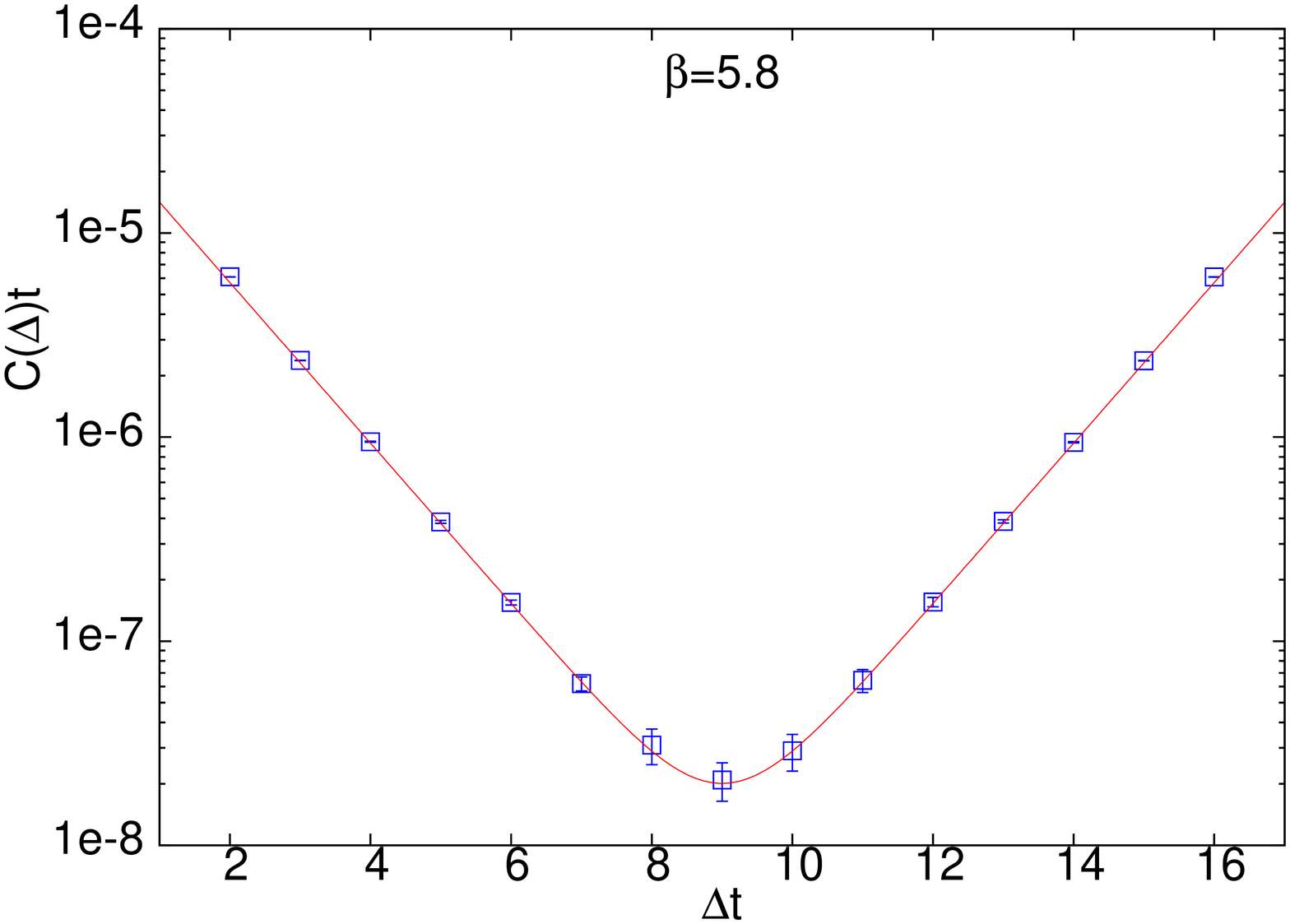}~~
\includegraphics[width=0.35\textwidth,angle=270]{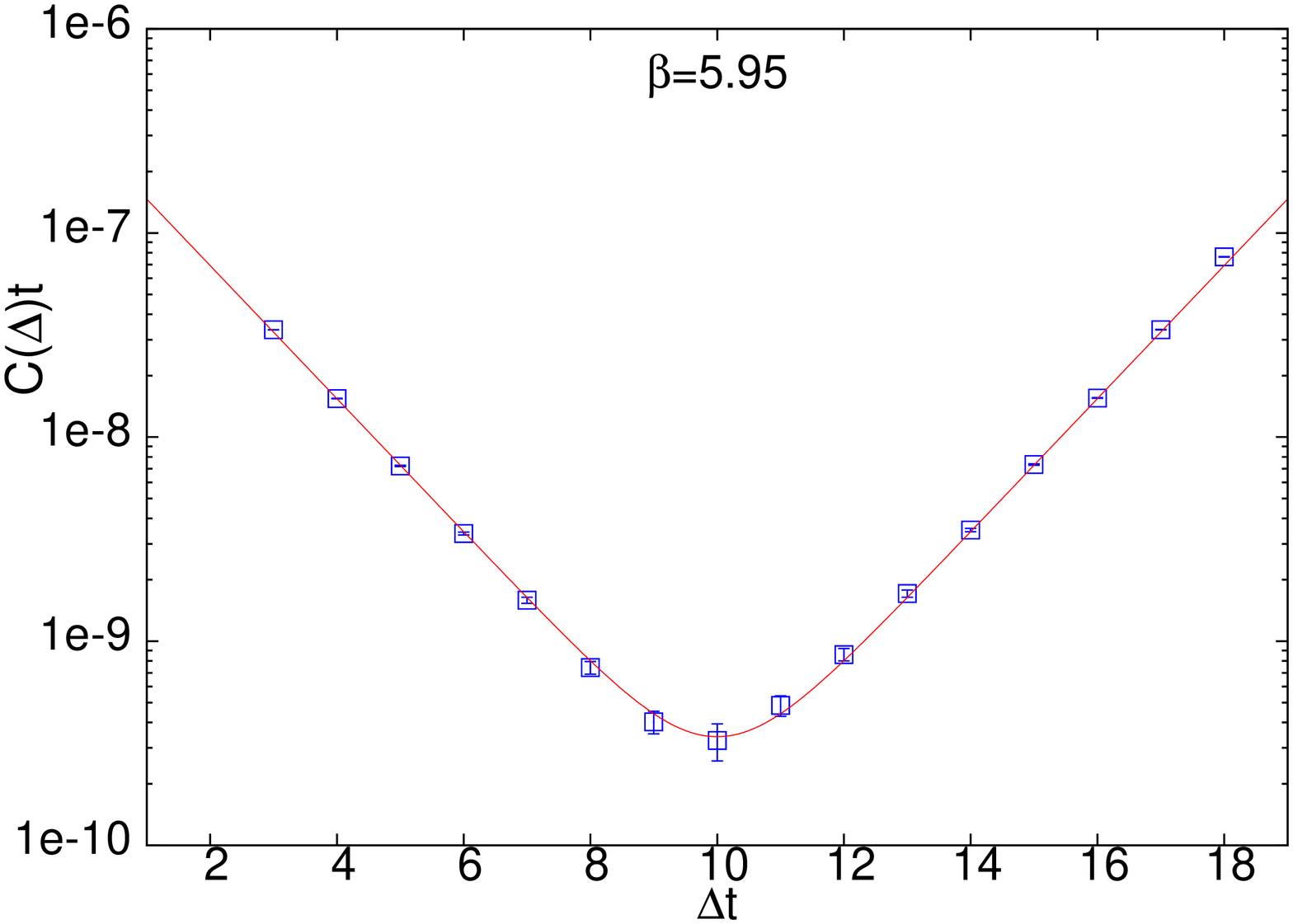}
\caption{Scalar correlators at $\beta$ = 5.8 (left) and 5.95 (right)}
\label{fig:corr-scalar}
\end{figure}

\begin{figure}
\includegraphics[width=0.35\textwidth,angle=270]{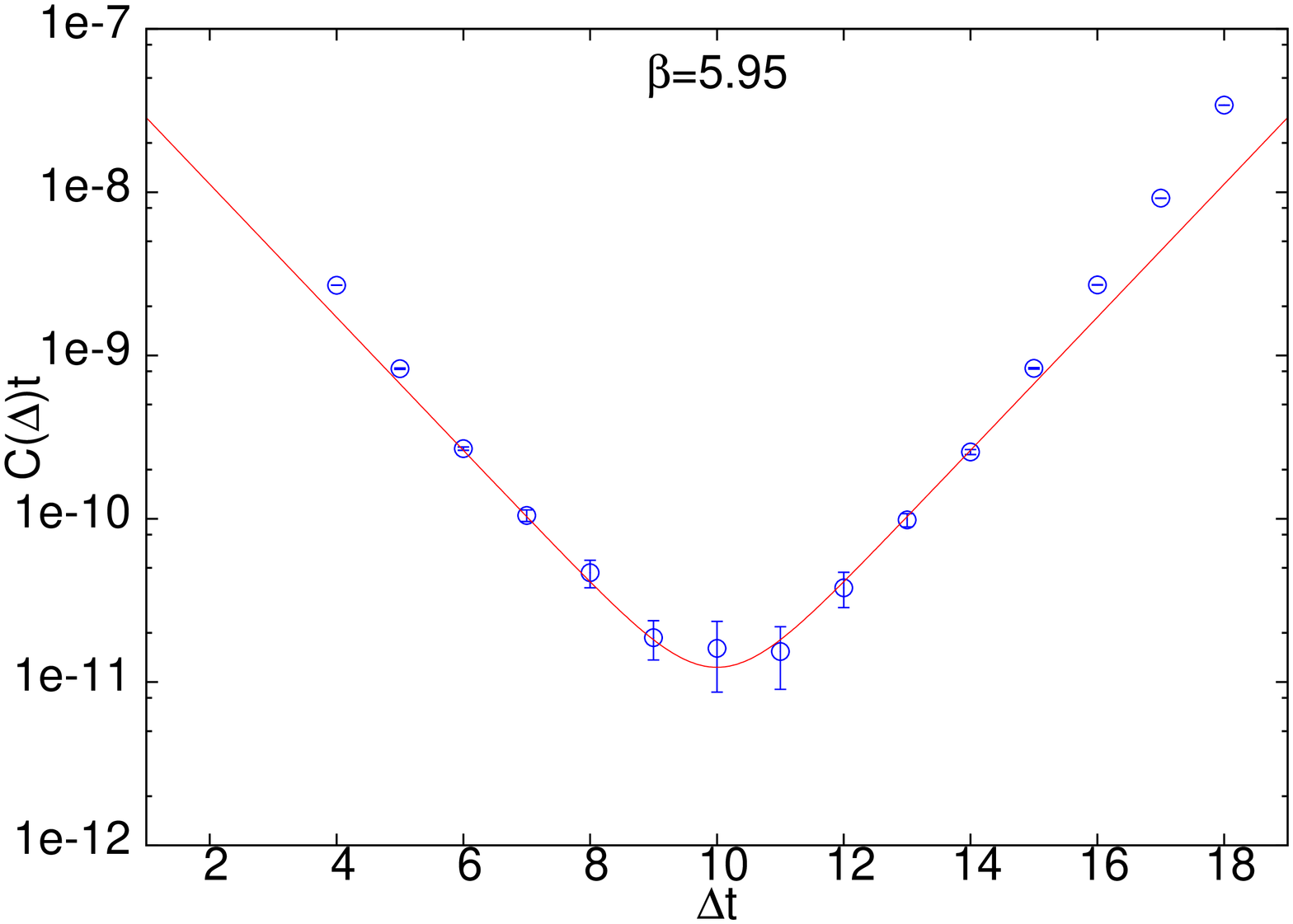}~~
\includegraphics[width=0.35\textwidth,angle=270]{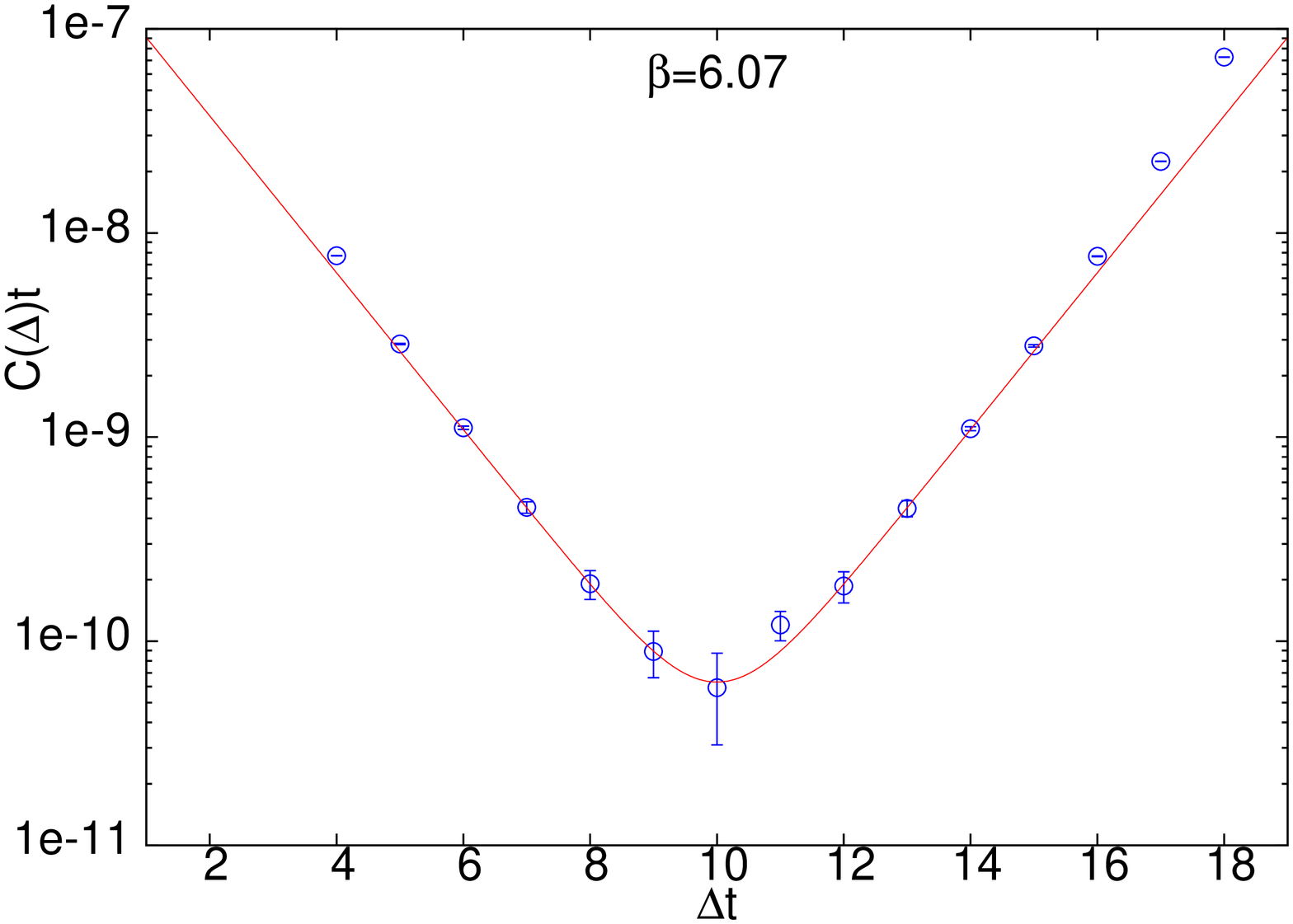}
\caption{Tensor correlators at $\beta$ = 5.95 (left) and 6.07 (right)}
\label{fig:corr-tensor}
\end{figure}

To extract masses from the correlators we fitted them to the form
\begin{equation}\label{fitform}
C(\Delta t)=A\left ( e^{-m\Delta t}+e^{-m(T-\Delta t)}\right )
\end{equation}
where $m$ is the glueball mass and $T$ is the full temporal extent of the lattice. Since the
correlator is symmetric about $T/2$, we fold the data and use only one half of the temporal range for the fits.
In figures \ref{fig:corr-scalar} and \ref{fig:corr-tensor} we show the scalar and tensor correlators along with 
the fitted correlators.

\begin{figure}
\includegraphics[width=0.35\textwidth,angle=270]{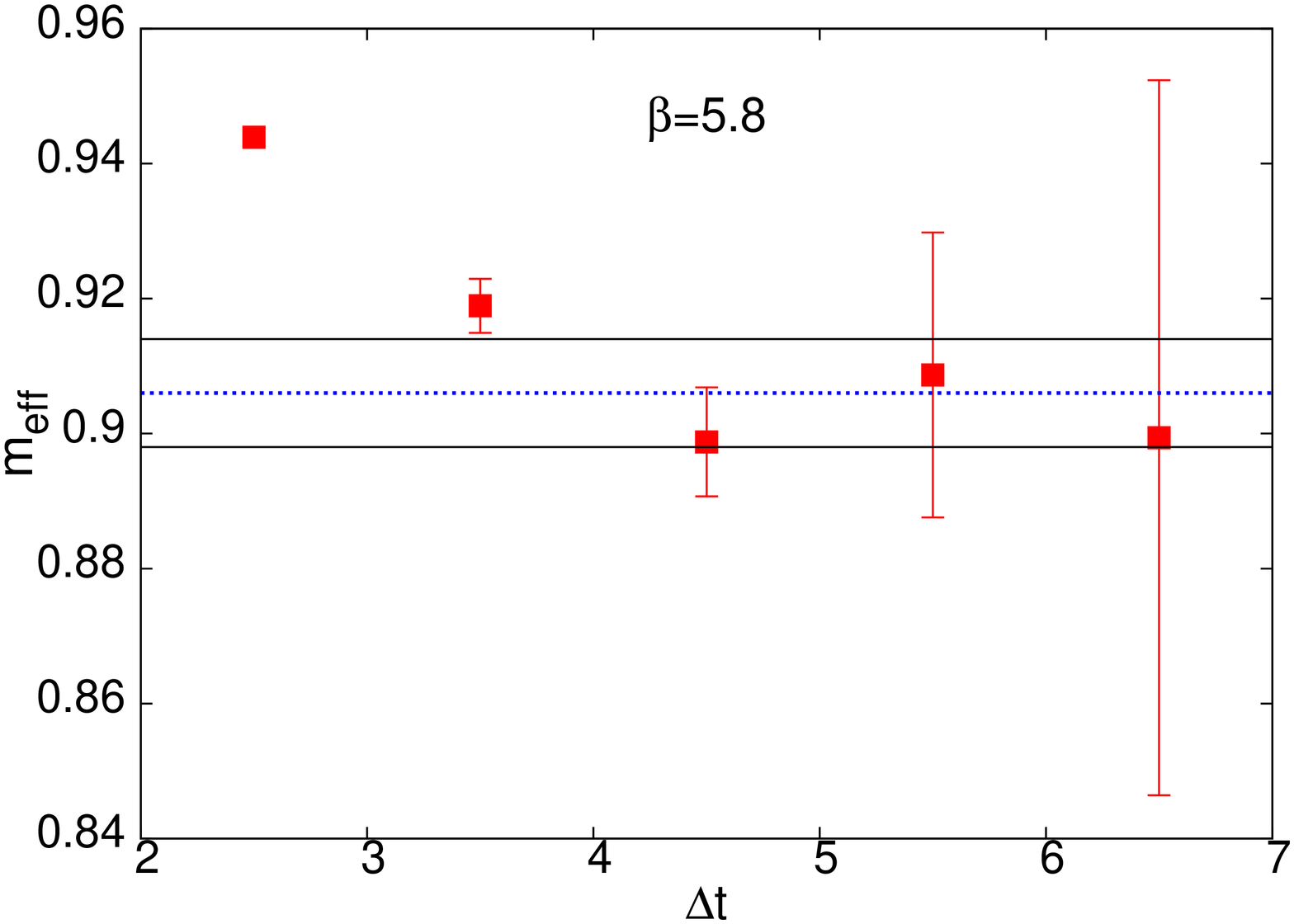}~~
\includegraphics[width=0.35\textwidth,angle=270]{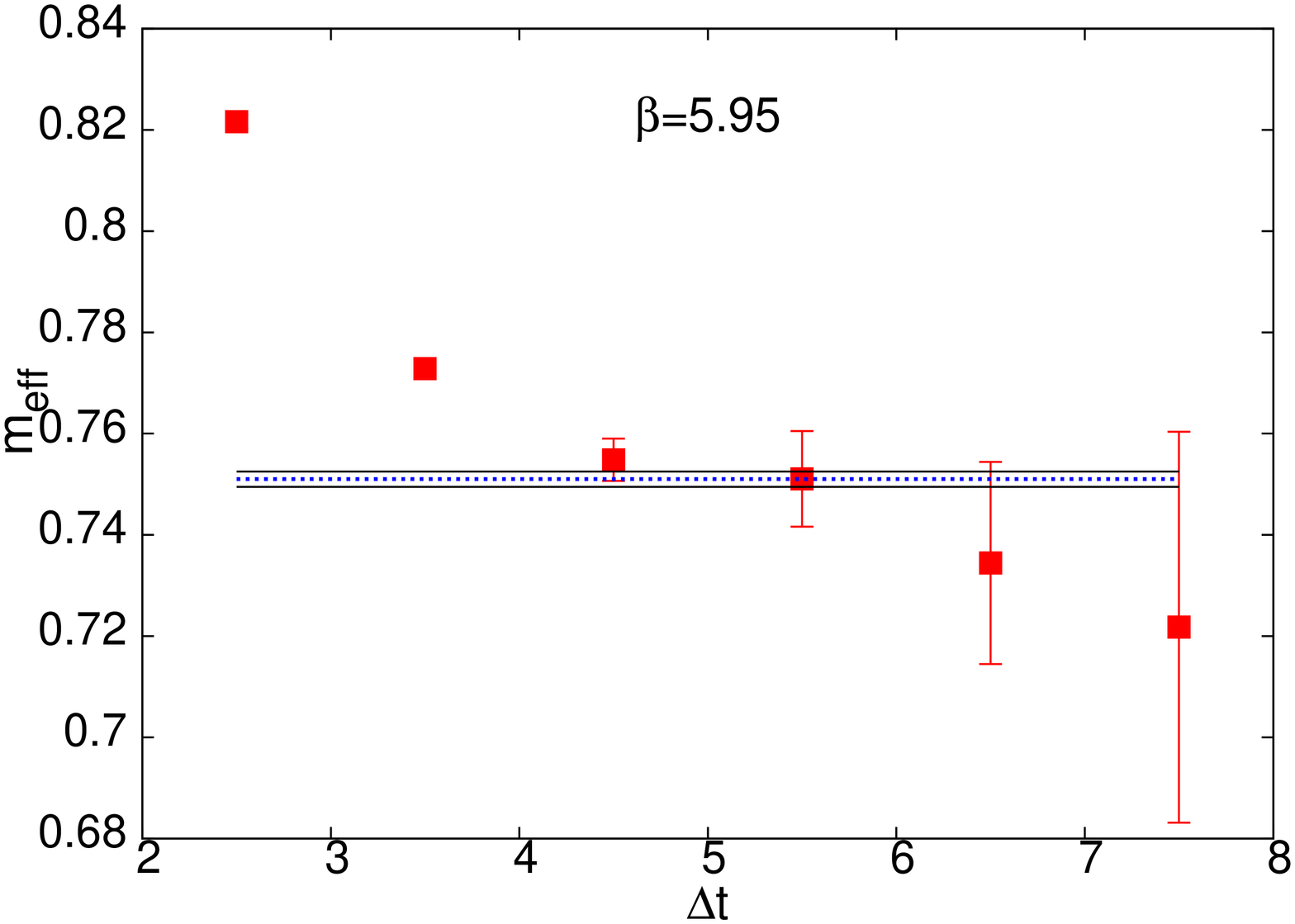}
\caption{Effective masses in the scalar channel at $\beta$ 5.8 (left) and 5.95 (right). The 
blue line is the mass obtained from the correlator fits. The two black lines show the error 
on the fitted mass.}
\label{fig:meff-scalar}
\end{figure}

\begin{figure}
\includegraphics[width=0.35\textwidth,angle=270]{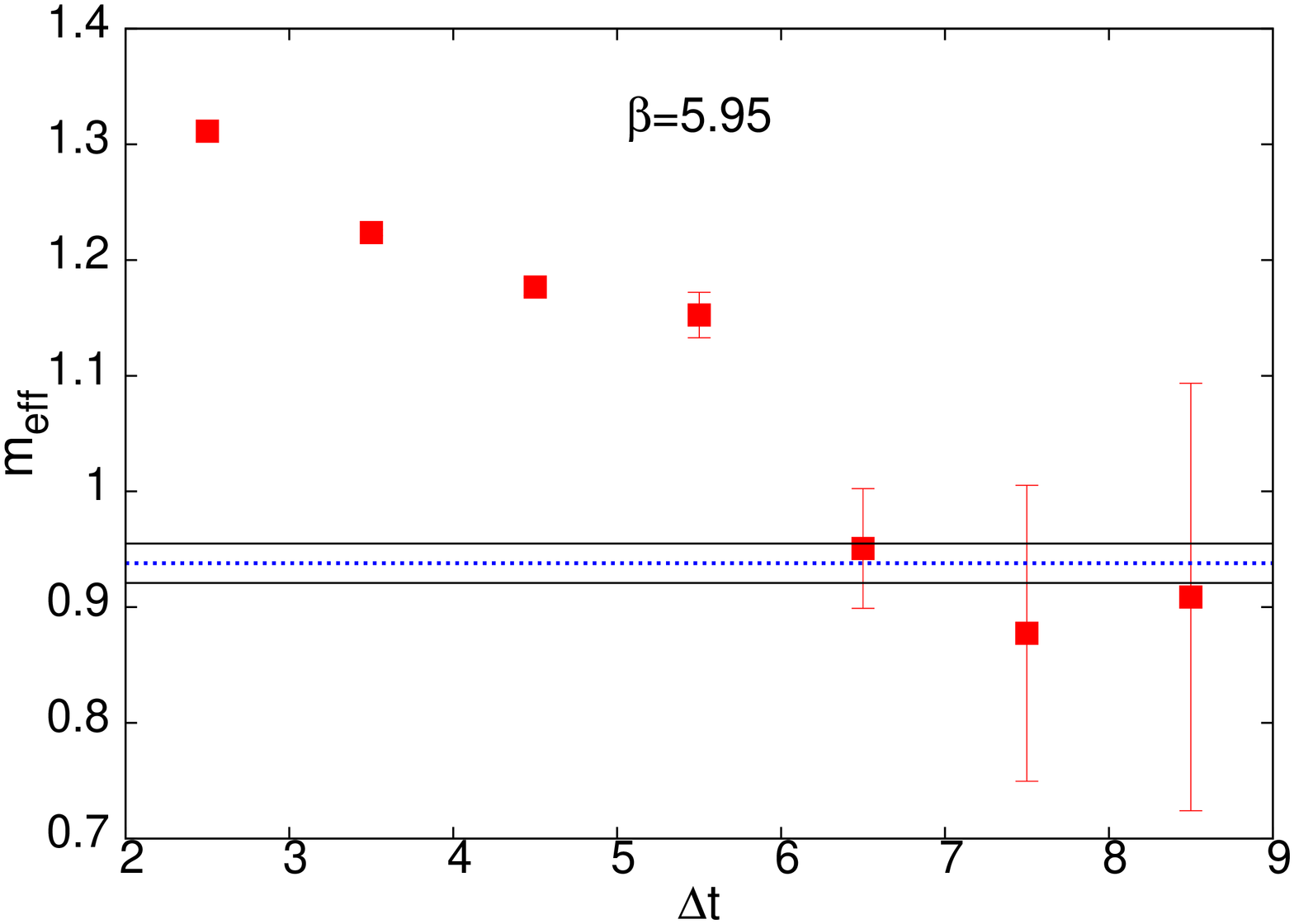}~~
\includegraphics[width=0.35\textwidth,angle=270]{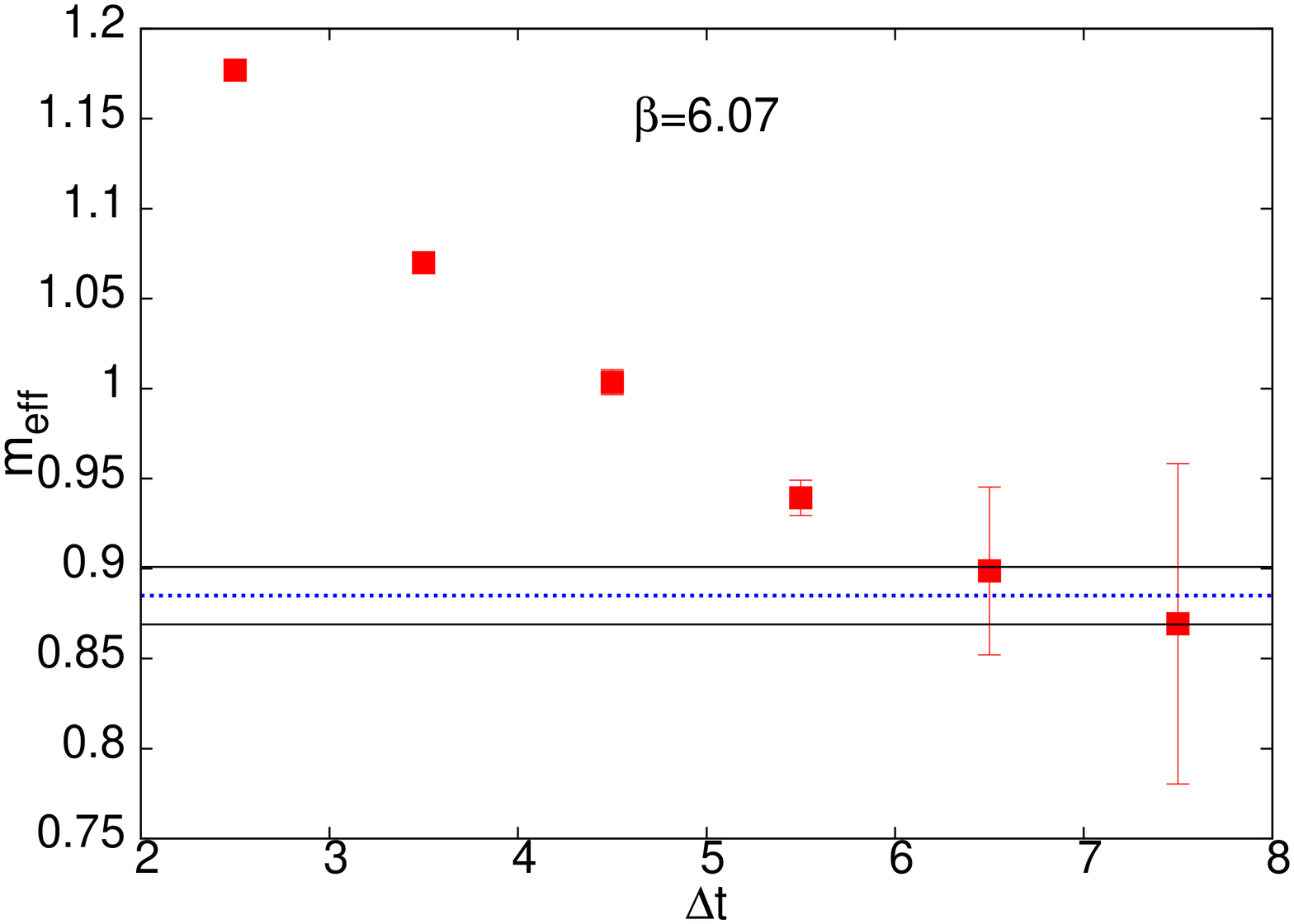}
\caption{Effective masses in the tensor channel at $\beta$ 5.95 (left) and 6.07 (right). 
Again the blue line is the mass from the correlator fits and the black lines show the error  
on the fitted mass.}
\label{fig:meff-tensor}
\end{figure}

In addition to the masses, we also compute the effective masses from the correlators as 
\begin{equation}\label{meff}
am_{\rm eff}=-\log\frac{\langle C(\Delta t +1)\rangle}{\langle C(\Delta t)\rangle}
\end{equation}
where $a$ is the lattice spacing. The effective masses are expected to decrease with 
increasing $\Delta t$ and finally settle to a stable value. This stable value is 
an estimate of the lightest mass in the concerned $J^{PC}$ channel. In figures 
\ref{fig:meff-scalar} and \ref{fig:meff-tensor} we plot the effective masses against 
$\Delta t$ in the scalar and tensor channels along with masses obtained from the 
correlator fits (blue line). We see that at the larger values of $\Delta t$, the 
effective masses match with the masses from the correlators.

As a cross-check of our results, we compare our data with that reported in reference \cite{Lucini}.
Our data is fully consistent with the results reported there and slightly more accurate. 

\begin{table}
\centerline{
\begin{tabular}{|c|c|c|c|c|} \hline
Lattice & $\beta$ & fit-range & $ma$ &${{\chi}^2}/d.o.f$  \\ \hline
${10^3}\times18$ & 5.7 & 5-9 & 0.952(11)& 0.066 \\ \hline
${12^3}\times18$ & 5.8 & 6-9 & 0.906(8)& 0.03 \\ \hline
${16^3}\times20$ & 5.95 & 5-10 & 0.7510(15)& 0.02 \\ \hline
\end{tabular}}
\caption{Global masses and fit parameters for the scalar channel}
\label{tab:fit-scalar}
\end{table}

\begin{table}
\centerline{
\begin{tabular}{|c|c|c|c|c|} \hline
Lattice & $\beta$ & fit-range & $ma$ &${{\chi}^2}/d.o.f$  \\ \hline
${12^3}\times18$ & 5.8 & 4-7 & 1.585(54)& 1.64 \\ \hline
${12^3}\times20$ & 5.95 & 6-10 & 0.938(17)& 0.12 \\ \hline
${12^3}\times20$ & 6.07 & 6-10 & 0.885(16)& 1.6 \\ \hline
\end{tabular}}
\caption{Global masses and fit parameters for the tensor channel}
\label{tab:fit-tensor}
\end{table}

To get an idea of the advantage of the current algorithm over the naive update algorithm, we did a few
runs for the same computer time using both the methods. For these runs we used the same $\beta$ values as 
our main run but smaller lattices. The results are reported in tables \ref{tab:perf-scalar} and 
\ref{tab:perf-tensor}. We see that depending on the channel and $\beta$, the gain in time is significant 
and can vary from 30 to more than 700. For the tensor channel at $\beta=6.07$ where we used a $6\times 6$ 
Wilson loop, we did not manage to get a signal using the naive method. Therefore we could not calculate 
the gain at that point.

\begin{table}
\centerline{
\begin{tabular}{|c|c|c|c|c|c|c|c|}\hline
Lattice & $\beta$ & sub-lattice & iupd & loop size & run-time (mins)& $\frac{{\rm error}_{naive}}{{\rm error}_{multilevel}}$ & gain (time)\\\hline
${10^3}\times18$& 5.7  & 3 & 30 & $2\times 2$ & 3850 & 5.7 & 32 \\ \hline
${6^3}\times18$ & 5.8  & 3 & 25 & $3\times 3$ & 1000 & 5.5 & 30 \\ \hline
${8^3}\times24$ & 5.95 & 4 & 50 & $5\times 5$ & 1100 & 18  & 324 \\ \hline
\end{tabular}}
\caption{Performance comparison for Scalar Channel}
\label{tab:perf-scalar}
\end{table}

\begin{table}
\centerline{
\begin{tabular}{|c|c|c|c|c|c|c|c|}\hline
Lattice & $\beta$ & sub-lattice & iupd & loop size & run-time (mins)& $\frac{{\rm error}_{naive}}{{\rm error}_{multilevel}}$ & gain (time)\\\hline
${6^3}\times18$ & 5.8  & 3 & 50  & $3\times 3$ & 12000 & 27 & 729 \\ \hline
${8^3}\times30$ & 5.95 & 5 & 100 & $5\times 5$ & 5775  & 20 & 400 \\ \hline
${10^3}\times30$& 6.07 & 6 & 130 & $6\times 6$ & 15000 & -  & - \\ \hline
\end{tabular}}
\caption{Performance comparison for tensor channel}
\label{tab:perf-tensor}
\end{table}

\section{Discussions}

The multilevel algorithm is very efficient for calculating quantities with very 
small expectation values. Operators in the tensor channel have zero expectation 
values and are therefore ideal for direct evaluation. For scalar operators we have 
subtracted the non-zero vacuum expectation values from the operators to get the 
connected correlators directly.

Correlation functions between large loops have the advantage that they have much 
less contamination from excited states compared to those between elementary 
plaquettes. Multilevel schemes allow us to estimate the expectation values
of the large loops with very high precision.

The efficiency of the algorithm depends crucially on choosing the optimal parameters 
for the algorithm such as the sub-lattice thickness and updates. These depend on $\beta$ 
quite strongly. In the range of $\beta$ we explored it seems that 0.5 fermi seems to be 
close to optimal for both the loop size and the thickness of the sub-lattice.

We observe that this error reduction technique works quite well at least in pure gauge 
theories. For a given computational cost, the improvement in the signal to noise ratio is 
several times to even a couple of orders of magnitude.

Finite volume effects is the largest source of systematic errors and to avoid them we choose 
our lattices such that $mL > 9$ \cite{MM}. For a more detailed discussion we refer the reader 
to reference \cite{PNS}.

\acknowledgments
The runs were carried out in portion on the cluster bought under the DST project SR/S2/HEP-35/2008
and in portion on the ILGTI part of the CRAY XE6-XK6 at IACS. The authors would like to thank DST, 
IACS and ILGTI for these facilities. The authors would also like to thank Peter Weisz for his 
comments on finite volume effects.

\end{document}